\documentclass[journal=ancac3,manuscript=article]{achemso}
\setkeys{acs}{maxauthors = 10}

\usepackage{textcomp}
\usepackage{hyperref}

\author{Alexander Markevich}
\affiliation[univie]
{University of Vienna, Faculty of Physics, Boltzmanngasse 5, 1090 Vienna, Austria}
\email{alexander.markevich@univie.ac.at}
\author{Bethany M Hudak}
\affiliation[NRL]{Naval Research Laboratory, Material Sciences and Technology, 4555 Overlook Ave SW, Washington, DC 20375, USA}
\author{Jacob Madsen}
\affiliation[univie]{University of Vienna, Faculty of Physics, Boltzmanngasse 5, 1090 Vienna, Austria}
\author{Jiaming Song}
\affiliation[NWU]{School of Physics, Northwest University, 1 Xuefu Avenue, Xi’an, Shaanxi 710127, China}
\author{Paul C Snijders}
\affiliation[ORNL1]{Materials Science and Technology Division, Oak Ridge National Laboratory, Oak Ridge, Tennessee 37830, USA}
\author{Andrew R Lupini}
\affiliation[ORNL2]{Center for Nanophase Materials Sciences, Oak Ridge National Laboratory, Oak Ridge, Tennessee 37830, USA}
\email{arl1000@ornl.gov}
\author{Toma Susi}
\affiliation[univie]{University of Vienna, Faculty of Physics, Boltzmanngasse 5, 1090 Vienna, Austria}
\email{toma.susi@univie.ac.at}

\title[]{Mechanism of electron-beam manipulation of single dopant atoms in silicon}

\begin{document}
  \clearpage
  \thispagestyle{empty}
\hspace{0pt}
\vfill
\noindent\textbf{DISTRIBUTION A}:  Approved for public release, distribution is unlimited.\\
\\
Notice: This manuscript has been authored by UT-Battelle, LLC, under Contract No. DE-AC0500OR22725 with the U.S. Department of Energy. The United States Government retains and the publisher, by accepting the article for publication, acknowledges that the United States Government retains a non-exclusive, paid-up, irrevocable, world-wide license to publish or reproduce the published form of this manuscript, or allow others to do so, for the United States Government purposes. The Department of Energy will provide public access to these results of federally sponsored research in accordance with the DOE Public Access Plan (http://energy.gov/downloads/doe-public-access-plan).
\vfill
\hspace{0pt}

\begin{tocentry}

\includegraphics[width=0.42\textwidth]{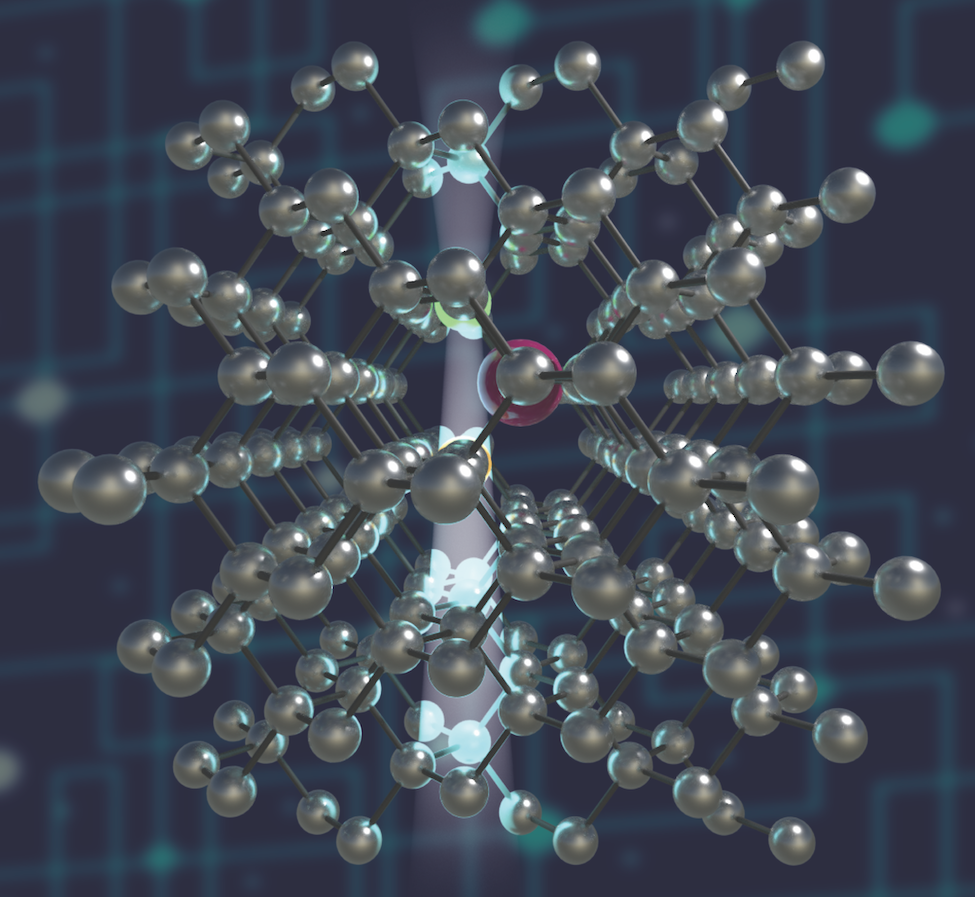}

\end{tocentry}

\begin{abstract}
The precise positioning of dopant atoms within bulk crystal lattices could enable novel applications in areas including solid-state sensing and quantum computation. Established scanning probe techniques are capable tools for the manipulation of surface atoms, but at a disadvantage due to their need to bring a physical tip into contact with the sample. This has prompted interest in electron-beam techniques, followed by the first proof-of-principle experiment of bismuth dopant manipulation in crystalline silicon. Here, we use first principles modeling to discover a novel \textit{indirect exchange} mechanism that allows electron impacts to non-destructively move dopants with atomic precision within the silicon lattice. However, this mechanism only works for the two heaviest group V donors with split-vacancy configurations, Bi and Sb. We verify our model by directly imaging these configurations for Bi, and by demonstrating that the promising nuclear spin qubit Sb can be manipulated using a focused electron beam.
\end{abstract}

\section{Introduction}
Quantum information processing has surged in recent decades from a speculative idea~\cite{benioff_computer_1980} into a technological race for quantum supremacy~\cite{arute_quantum_2019}. Multiple viable architectures with various strengths and weaknesses have been proposed~\cite{divincenzo_physical_2000}, including solid-state qubits. Due to our reliance and technological capabilities in silicon manufacturing, the nuclear spins of positively charged donor atoms within crystalline silicon have attracted particular interest~\cite{kane_silicon-based_1998,morton_embracing_2011}. However, two factors make such qubits challenging: the precise placement of dopants into an ordered array, and their individual addressing for quantum logic operations. The mere detection of single atoms inside a semiconductor is extremely difficult, and there are only a few techniques capable of this feat---particularly, atom probe tomography~\cite{kelly_atom_2007} and aberration-corrected scanning transmission electron microscopy~\cite{nellist_direct_2004} (STEM).

Phosphorus ($^{31}$P) as a dopant has attracted the greatest interest due to its long nuclear spin coherence lifetime, enabling it to function as a quantum memory~\cite{morton_solid-state_2008}. Typically, traditional non-deterministic ion implantation and post-hoc lithography have been used for fabricating devices~\cite{pla_single-atom_2012}. A degree of deterministic placement control has been enabled by scanning tunneling microscopy based hydrogen depassivation lithography of Si surfaces~\cite{randall_atomic_2009}, followed by molecular beam epitaxy growth to complete the crystal over a chemically introduced donor~\cite{fuechsle_single-atom_2012}. However, the spin dephasing time for the hyperfine-coupled electron on P sites is rather short~\cite{muhonen_storing_2014}, hindering its electric-field control. Very recently, antimony (Sb) has been suggested as a promising alternative, with two orders of magnitude higher dephasing time and robust electrical control via its nuclear quadrupole moment~\cite{asaad_coherent_2020}.

Despite some advances, dopant placement remains the major hurdle for further progress. Electron-beam manipulation of covalently bound lattice impurities was recently proposed as a new kind of atomically precise manipulation tool~\cite{susi_siliconcarbon_2014,kalinin_fire_2016}. Initially, this was limited to impurities in carbon nanomaterials including graphene~\cite{susi_manipulating_2017,dyck_placing_2017,tripathi_electron-beam_2018,su_engineering_2019} and single-walled carbon nanotubes~\cite{mustonen_electron-beam_2019}, but a similar capability was recently also demonstrated for heavy Bi dopants in silicon~\cite{hudak_directed_2018,jesse_direct_2018}. However, the atomistic mechanism for their movement within the crystal was not clear, nor was it known whether other dopants could be affected, which beam energies are optimal, nor whether the surrounding lattice is irreparably damaged.

The electron-beam energy in a (S)TEM is typically in the range of 30–-300\,keV, while displacement threshold energies are on the order of 10\,eV. However, conservation of momentum implies that only a very small fraction of the kinetic energy can be transferred from the fast electron to a nucleus in a single elastic collision. Electronic interactions or multiple collisions might also contribute depending on the material and imaging conditions. In an insulating sample, ionization or “radiolysis” would complicate the picture~\cite{susi_quantifying_2019}, but the electrical conductivity of highly doped Si should mitigate such damage. Moreover, in a recent study of atomic displacements in MoS$_2$ under electron irradiation,~\cite{kretschmer_formation_2020} it was demonstrated that electronic excitations strongly affected displacement threshold energies at lower accelerating voltages but made only a minor contribution at a primary beam energy of 80\,kV. The relatively low beam current typical for aberration-corrected STEM suggests that sample heating is negligible because of the thermal conductivity of the sample~\cite{borner_evidence_2016}. The dominant beam interaction mechanism in these samples is therefore expected to be elastic collisions transferring between 10--20\,eV from single fast incident electrons to individual nuclei. The knock-on threshold energy for bulk Si has been calculated~\cite{holmstrom_threshold_2008} at about 12.5\,eV, corresponding to an electron beam energy of about 140\,keV.

In this study, we combine density functional theory molecular dynamics (DFT/MD) and aberration-corrected STEM to systematically study the behaviour of group V dopants in silicon under electron irradiation. Similar to heteroatom-doped carbon systems~\cite{susi_atomistic_2012,susi_siliconcarbon_2014,su_engineering_2019}, momentum-conserving electron scattering at modest primary beam energies can only affect the lighter Si neighbors of the impurity atom, allowing its movement to be controlled by directing the electron beam at a chosen lattice site. Since any beam-induced dynamics occur on timescales that are far below the experimental time resolution~\cite{susi_siliconcarbon_2014}, modeling is required to understand the atomistic details of the process.

We reveal that dopant manipulation in crystalline silicon proceeds via a novel type of nondestructive mechanism we call \emph{indirect exchange}: the impacted Si is knocked into a metastable interstitial configuration, with the impurity taking its original lattice position in the subsequent recombination. Contrary to the process in graphene, the primary knock-on atom does not end up as a neighbor of the impurity, but rather as a next-nearest neighbor, displacing another Si atom during the dynamics.

Our simulations show that this mechanism only works with the heavier two pnictogen dopants, Sb and Bi, which induce greater lattice distortion, and whose split-vacancy configuration in the case of Bi we are occasionally able to image directly, whereas for the lighter two, As and P, the ejected Si interstitial  recombines with a vacancy without exchange of atom positions. To confirm these predictions, we demonstrate for the first time the manipulation of Sb impurities in a thin crystalline slab of silicon, finding that similarly to Bi these can be successfully manipulated using focused electron irradiation.


\section{Methods}

\subsection{Computational Methods}

\subsubsection{DFT simulations}
All simulations were performed using density functional theory (DFT) as implemented in the GPAW package~\cite{enkovaara_electronic_2010}. We used the PBE exchange-correlation functional~\cite{perdew_generalized_1996}, a localized (\textit{dzp}) basis set~\cite{larsen_localized_2009}, and a real-space grid spacing of 0.2\,\AA. A cubic supercell of 216 atoms was used in all calculations. The Brillouin zone was sampled using a $3\times3\times3$ \textbf{k}-point mesh according to the Monkhorst-Pack scheme~\cite{monkhorst_special_1976}.

It should be noted that while manipulation of dopants has been performed in Si slabs of $\sim$10-–20\,nm thickness, for simulations we used a periodic supercell representing an infinite crystal. In a sub-surface layer of a real sample, lattice strains induced by the surface layer reconstruction can affect the dynamic behavior of crystal defects. Considerable lattice strain has been shown to extend for five atomic layers into the crystal lattice,~\cite{appelbaum_theory_1978,kamiyama_ab_2012} which in the case of Si is $\sim$0.5 nm. In our experiments on dopant manipulation, we have estimated typical depths of the dopant atoms that have been probed at about 3-–5\,nm from the surface.~\cite{hudak_directed_2018} At such a distance from the surface, the lattice strain induced by the surface reconstruction should be negligible and the phenomena observed experimentally are characteristic for a bulk crystal volume. 

To simulate the dynamics of the atoms induced by electron irradiation, we used Velocity-Verlet (NVE) molecular dynamics~\cite{swope_computer_1982} with a time step of 0.5\,fs. The effect of momentum-conserving electron scattering on Si neighbours of a dopant was modelled by assigning a momentum that corresponds to a specific transferred energy to one Si atom. Threshold energies were estimated by varying the amount of energy transferred to the Si atom at 0.1\,eV intervals. A detailed description of the methodology can be found in Ref.~\citenum{susi_isotope_2016}.

Heating of systems with As and P dopants and interstitial Si was performed using Langevin (NVT) dynamics~\cite{schneider_molecular-dynamics_1978} up to 1200\,K or until any structural changes occurred. Random snapshots of the system were then taken at temperatures in the range of 800--1200\,K and six Velocity-Verlet simulations were performed for each dopant until the recombination of the defect occurred. The energy barrier for the recombination of the Bi-V complex with a Si interstitial was calculated with the climbing-image nudged elastic band (cNEB) method~\cite{henkelman_climbing_2000} using nine system images with a spring constant of 0.1\,eV/\AA, relaxed until all atomic forces were below 0.02\,eV/\AA.

\subsubsection{Image simulations}
All image simulations were performed using the multislice code \textsc{abTEM}~\cite{madsen_abtem_2021}. Based on the experimental HAADF contrast, we assumed a slab thickness of 15\,nm and that the dopants were near the center of the slab. The size of the structure prohibits fully optimizing all the positions at the \emph{ab initio} level of theory; hence, smaller supercells around the defects were optimized and inserted into an idealized slab. The acceleration voltage, beam convergence semi-angle, and detector semi-angles were set to the experimental values. The remaining unknown imaging parameters were estimated by matching the image contrast through a grid search. The optimal parameters were a defocus of 180\,\AA, a spherical aberration of 40\,{\textmu}m, beam tilt in $x$ and $y$ of 13 and 9\,mrad, and a source spread of 0.8\,\AA\ (FWHM). We used the PRISM algorithm~\cite{ophus_fast_2017} to quickly simulate a large number of required images for the grid search. The final image was simulated using the standard multislice algorithm with thermal diffuse scattering included in the frozen phonon approximation.


\subsection{Experimental Materials and Methods}

\subsubsection{Sample growth}
Sb-doped Si films were grown on Si(100) substrates. The Si(100) surface was prepared with a (2$\times$1) reconstruction using standard ultra-high-vacuum degassing and flashing procedures, and surface quality examined by scanning tunneling microscopy and low-energy electron diffraction. On the Si(100)-2$\times$1 surface, approximately 0.5 monolayers (ML) of Sb atoms was subsequently deposited with the substrate held at 550\,{\textdegree}C, using a standard effusion cell held at 400\,{\textdegree}C. This Sb dose, higher than desirable for qubit applications, was chosen in order to have a sufficient doping concentration for good visibility and easy dopant manipulation. Next, a nominally $\sim$5 ML Si capping layer was grown with the substrate held at room temperature and annealed at 450\,{\textdegree}C for 5\,s. A final capping layer of 20~ML Si was grown with the substrate held at room temperature and annealed at 720\,{\textdegree}C for 1\,s to crystallize.

\subsubsection{Specimen preparation}
Cross-sectional STEM samples were prepared by taking the as-grown sample, cutting it in half to produce a sandwich, and then slicing it into sections that were glued to a support grid. The samples were mechanically polished to approximately 10\,{\textmu}m, then ion-milled to electron transparency. Immediately before imaging, the samples were baked at a nominal 140\,{\textdegree}C in vacuum overnight to reduce surface contamination. Thickness for typical images was measured by electron energy-loss spectroscopy as about 0.1--0.2 mean free paths (about 10--20\,nm), although both thicker and thinner regions were examined. Images taken far away from the film and glue indicate the existence of an amorphous oxide layer about 1\,nm in thickness. After ion-milling, the Si surface was chemically and structurally disordered. We found that there was often apparent motion in the images, due to the surface oxides and overlying amorphous carbon. Repeated intense illumination of the same area usually resulted in milling of the surface Si.

\subsubsection{Scanning transmission electron microscopy}
Aberration-corrected STEM was performed in a Nion UltraSTEM 200 operated at 160\,keV at a low beam current of 20$\pm$2\,pA and a nominal beam convergence semiangle of 30\,mrad. We acquired images with a high-angle annular dark-field (HAADF) detector with a nominal semiangular range of 60--400\,mrad under a variety of dose conditions, including field of view and scan speed. An ImageJ smoothing filter that replaces each pixel with the average of its 3$\times$3 neighbors was applied to Fig.~\ref{fig:manipulation} to enhance contrast. Sb impurities are identified by their characteristic $Z$-contrast, and their presence in the sample was also verified by electron energy loss spectroscopy collected at 200\,keV with a dispersion of 0.5\,eV/ch and an acquisition time of 1\,s/pixel. The additional minor EEL intensity in Fig.~\ref{fig:manipulation} starting at $\sim$766\,eV likely corresponds to the Sb $M_3$ edge, while the more intense features starting at $\sim$931\,eV can likely be attributed to the $L_{2,3}$ edge of Cu contamination near the disordered surface layer.

\section{Results and discussion}

\subsection*{Uncovering the manipulation mechanism}
To uncover the mechanism of directed dopant movement in silicon, we performed DFT/MD simulations to model the effect of momentum-conserving elastic electron scattering (ie. knock-on impacts) on the lattice atoms. We distinguish two types of lattice jumps for the impurities: either diagonally or laterally between the atomic columns as viewed along the [110] zone axis. While such jumps are equivalent from the perspective of lattice symmetry, we observe slight differences in terms of their electron-beam induced dynamics. Similar to the case of graphene, directed manipulation is made possible by the fact that beam-induced dynamics are triggered by an electron impact on one of the lattice atoms neighboring the impurity.

Let us first consider the case of the heaviest stable group V dopant, bismuth. In the original experimental work on Bi manipulation in crystalline silicon, the authors used electron energies of 60, 100, 160 and 200\,keV~\cite{hudak_directed_2018,jesse_direct_2018}. For 60 and 100\,keV, almost no changes in the structure were observed, while for 160\,keV, the positioning of dopants was possible by rapidly dragging a stationary electron probe over multiple atomic columns. This clearly indicated that a knock-on process was involved; indeed, originally the movement was thought to occur via the sputtering of Si atoms by the beam, followed by impurity migration via the created vacancies. As we show below, the actual mechanism is somewhat more complicated---but also more favorable for electron-beam manipulation.

\begin{figure}[t]
\centering
\includegraphics[width=\textwidth,keepaspectratio]{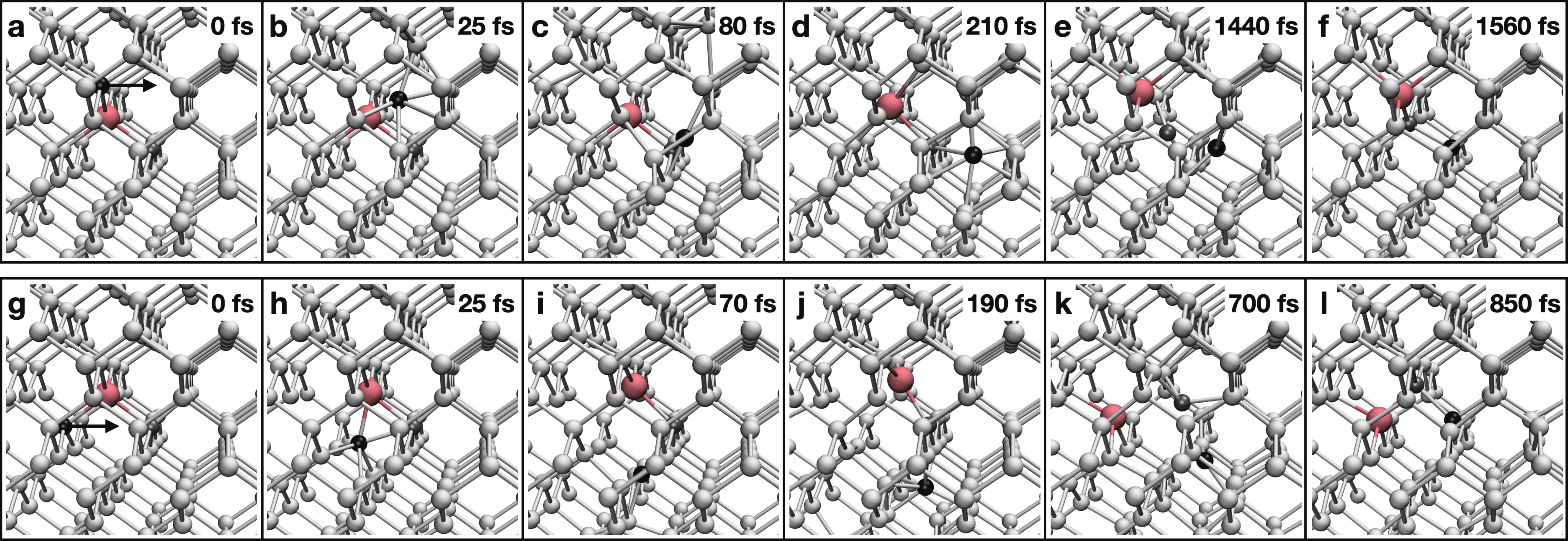}
\caption{\textbf{Indirect exchange of Bi and Si lattice positions induced by an electron impact.} Snapshots from DFT/MD trajectories showing the jump of a Bi impurity, shown in pink, by one lattice site through the indirect exchange mechanism. The Si atom shown in black has been assigned a momentum corresponding to a kinetic energy of 14.5\,eV in the [110] lattice direction as indicated by the arrow. The Bi impurity is shown in pink color and Si atoms within the bulk in light grey. We distinguish two types of impurity jumps: either diagonally (a-f) or laterally (g-l) between the atomic columns as viewed along the [110] zone axis of the experimental geometry (here, the dynamics are visualized from the side for clarity). During the dynamics, the Bi moves to the lattice position originally occupied by the impacted Si, while the latter becomes its second-nearest neighbour upon recombination, pushing the Si neighbor of the impurity shown in dark grey into the lattice site vacated by the dopant. (See also Supporting Information.)}
\label{fig:trajectory}
\end{figure}

For 160\,keV primary beam energy, the maximum kinetic energy that an elastically back-scattering electron can transfer to a stationary Si atom is $\sim$14.5\,eV, with the transferred momentum directed along the beam direction. Figure~\ref{fig:trajectory}a-f shows snapshots from a DFT/MD trajectory where a diagonal Si neighbour of Bi is assigned a velocity that corresponds to 14.5\,eV of kinetic energy in the [110] direction. The impacted Si atom moves to an interstitial configuration (Fig.~\ref{fig:trajectory}d), leaving behind a bismuth-vacancy (Bi-V) defect. This configuration is metastable, with the impurity atom further taking over the lattice position originally occupied by the impacted Si atom, while the latter becomes its second-nearest neighbour, displacing another Si atom to fill the vacant lattice site. In many of our MD trajectories, this occurred within the simulation time of 1--2\,ps. We further estimate a nudged elastic band (NEB) energy barrier of only $\sim$70\,meV for the recombination.

Notably, no atoms are---nor need to be---sputtered from the structure. Thus, this process is similar to the {\it direct exchange} mechanism discovered earlier for substitutional dopants in graphene~\cite{susi_siliconcarbon_2014,su_engineering_2019}, where the impurity atom exchanges places with a neighboring lattice C atom. However, in the case of bulk silicon, the impacted Si neighbour of the impurity instead ends up as its second-nearest neighbour, and thus we call this process {\it indirect exchange}.

The dynamics discussed thus far result in the Bi dopant moving diagonally between the columns as viewed along the [110] zone axis, and such lattice dynamics occur for transferred energies above 13.6\,eV. Below this energy, the impacted Si atom does not reach the interstitial configuration and quickly returns to its initial position. Above $\sim$22\,eV, the dynamics becomes more complex, with multiple atoms displaced from their lattice sites.

For lateral Bi jumps across the columns (Fig.~\ref{fig:trajectory}g-l), the dopant has two Si neighbours that lie in the same atomic column along the probe direction. Our simulations show that only impacts on the Si positioned closer to the probe and thus before the plane of the impurity can result in the movement of the dopant between the columns through the indirect exchange mechanism. The calculated threshold energy for this process is 13.9\,eV, slightly higher than that for jumps between the columns.

It should be noted that lateral jumps also result in the change of the dopant position perpendicular to the (110) plane by one lattice site. This provides the possibility to control the depth of a dopant within the slab by repeated lateral jumps, but only in the direction towards the electron source (although measuring this is challenging). Conversely, manipulation strictly within the same plane is only possible for diagonal jumps.

\subsection*{Differences amongst group V dopants}
To explore the potential of the electron-manipulation technique further, we systematically modelled the remaining group V dopants in silicon. Our simulations show that Sb impurities can be manipulated in the same way as Bi, with lattice jumps occurring through the indirect exchange mechanism. Threshold energies for diagonal and lateral Sb movement were calculated to be 13.0\,eV and 13.6\,eV, respectively. However, in the case of the lighter impurities, As and P, no indirect exchange process was observed in our simulations. Instead, in some of the trajectories where the impacted Si atom moved to an interstitial configuration, recombination occurred with no change in the dopant's lattice position. We further performed simulations at elevated temperatures in the range of 800--1200\,K with a starting configuration where the Si atom is at the interstitial site. In all simulations, the recombination occurred without the movement of the dopant atom to the target lattice site.

\begin{figure}
\centering
\includegraphics[width=0.8\textwidth,keepaspectratio]{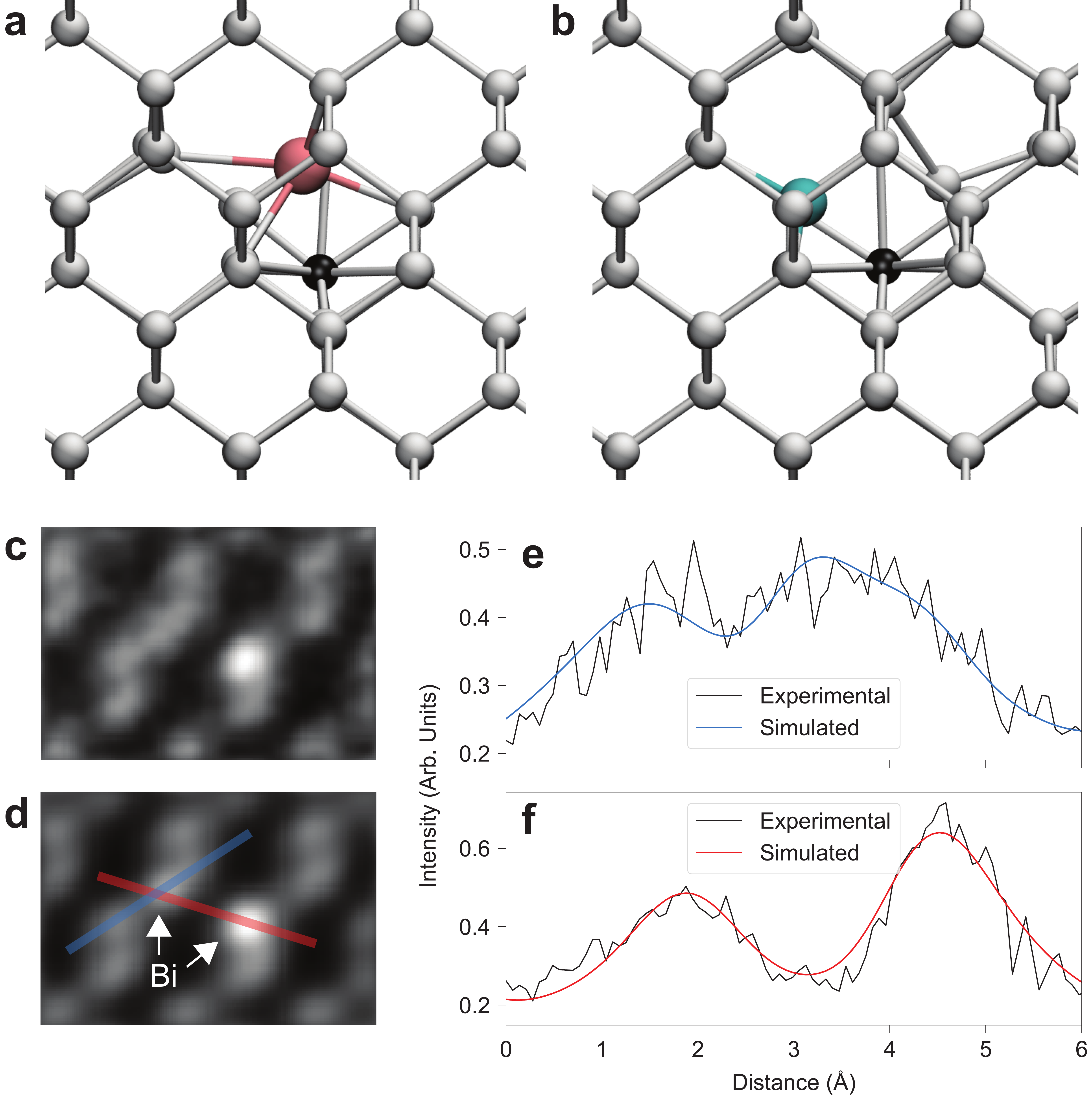}
\caption{\textbf{Group V dopant-vacancy complexes in Si.} (a,b) Atomic structures of dopant-vacancy complexes with a Si interstitial optimised with DFT for (a) heavier Bi with a split-vacancy configuration (Sb appears similar), and for (b) lighter As with substitutional dopant-vacancy configuration (P appears similar). (c) HAADF image (Gaussian filtered) and (d) image simulation of a Bi split-vacancy near a Bi substitution in Si(110) at 160\,keV. The overlaid line profiles indicate the intensity plotted over (e) the Si column, Bi split vacancy, and Si column direction (red) and (f) the Bi split vacancy and Si column with Bi substitution direction (blue), with an excellent quantitative match obtained. (See also Supporting Information.)}
\label{fig:interstitial}
\end{figure}

Such a clear difference in the behaviour of larger and smaller dopant atoms can be explained by the atomic structure of their vacancy complexes. Oversized impurities, such as Bi and Sb, form a split-vacancy structure, whereas As and P adopt a substitutional dopant-vacancy geometry instead~\cite{hohler_vacancy_2005}. Similar structures are formed during the indirect exchange process. When a Si neighbour is knocked into an interstitial configuration, Bi and Sb dopants move towards the newly vacant lattice site (Fig.~\ref{fig:interstitial}a), opening an easy route for recombination, where only one additional lattice Si atom needs to be displaced. In contrast, As and P remain close to their original lattice positions even when their neighbor is displaced (Fig.~\ref{fig:interstitial}b). MD simulations at elevated temperatures suggest that in this case, the recombination of the interstitial Si, which involves the displacement of two additional lattice atoms, turns out to be energetically more favourable than the jump of the dopant.

Remarkably, in the Bi-doped sample, we were occasionally able to record intense atomic contrast between the diagonal Si columns, far too high to correspond to a Si atom, which, even if stationary, would not be visible compared to the lattice contrast. However, this atomic contrast is consistent with a single Bi atom as verified by quantitative image simulations (Fig.~\ref{fig:interstitial}c-d). We thus believe that the impacted Si interstitial can occasionally diffuse away from the site instead of recombining, allowing the split-vacancy configuration to be directly recorded. This idea is supported by the fact that such contrast was observed more frequently at 200\,keV, where the electron beam can transfer more energy to the impacted Si atom, facilitating its escape.

\subsection*{Electron-beam manipulation of antimony}
\begin{figure}
\centering
\includegraphics[width=0.9\textwidth,keepaspectratio]{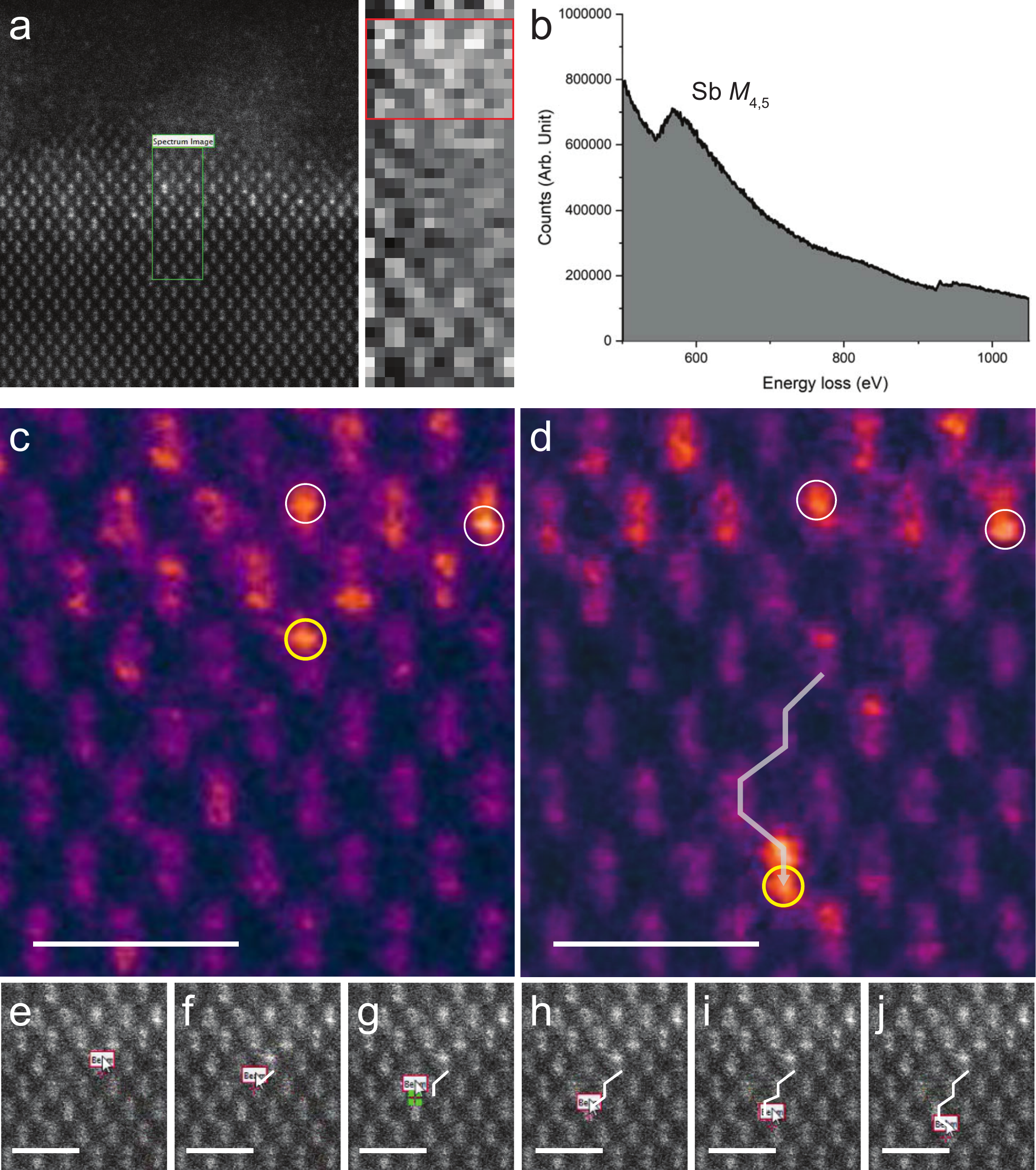}
\caption{\textbf{Antimony manipulation.} Identification and electron-beam manipulation of Sb dopants in Si(110). (a) HAADF image of the Sb-doped Si sample acquired at 200\,keV, with a green overlaid rectangle corresponding to the electron energy-loss spectrum image shown in the middle. The spectrum in (b) was summed over the 15$\times$10 pixels indicated by the red rectangle to enhance the signal to noise ratio, showing a prominent Sb $M_{4,5}$ edge starting at $\sim$528\,eV (see Methods). (c) HAADF close-up of the region at the lower end of the green rectangle in (a) captured before manipulation at 160\,keV. Yellow circles indicate the targeted Sb atom, and white circles mark Sb atoms that remain stationary during the process. (b) Image acquired after direct Sb positioning, with the manipulation path traced as a white line. (e-j) Screen-captured frames showing the position of the electron beam as it was used to guide the Sb dopant. Scale bar is 1\,nm. (See also Supporting Information.)}\label{fig:manipulation}
\end{figure}

To confirm our theoretical predictions, we attempted to experimentally manipulate Sb impurities using the focused electron probe of an aberration-corrected Nion UltraSTEM200 operating at 160\,keV. By 'dragging' the beam along atomic columns adjacent to the Sb-containing column, we can controllably direct the position of the Sb dopant in the same manner that Bi atoms were directed~\cite{hudak_directed_2018}. Since the beam cannot be simultaneously scanned and manually directed, an image (Fig.~\ref{fig:manipulation}c) is acquired and used as a survey image. The beam is manually positioned over a series of neighboring atomic columns in the survey image, and then a second scanned image is acquired (Fig.~\ref{fig:manipulation}d), revealing the new location of the Sb atom. Fig.~\ref{fig:manipulation}e-j illustrate the path traced by the beam. As in the case of Bi~\cite{hudak_directed_2018}, the Sb has moved several columns away from its initial position, following the path that the operator directed with the electron beam. (The additional bright contrast near the endpoint is likely due to dopants at different depths being simultaneously dragged into position, which is a challenge for 3D manipulation.) White circles in Fig.~\ref{fig:manipulation} indicate Sb atoms that remain stationary during the process, highlighting that the movement is not simply an overall drift. Thus, we are able to guide the position of the Sb dopant at 160\,keV, in agreement with our DFT/MD prediction.

\subsection*{Influence of the atom emission angle}
\begin{figure}[hb!]
\centering
\includegraphics[width=0.75\textwidth,keepaspectratio]{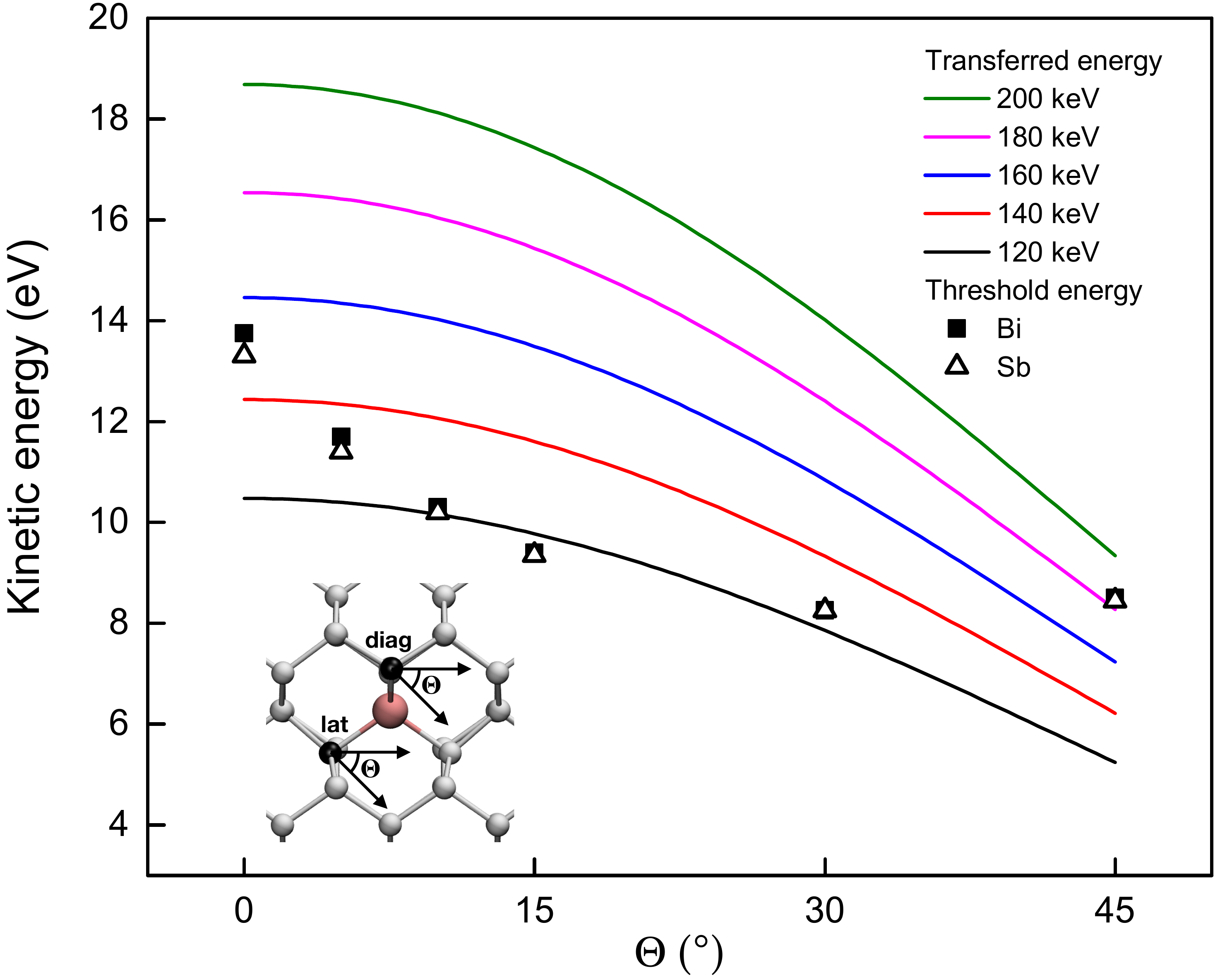}
\caption{\textbf{Manipulation energetics.} Calculated threshold energies for the indirect exchange for Bi (black squares) and Sb (open triangles) dopants (average of values for diagonal and lateral jumps (see Table~\ref{tab:thresholds}), as a function of the impacted Si atom (colored black in the inset) emission angle $\Theta$. Solid lines show the maximum kinetic energy that can be transferred to a stationary Si atom in the corresponding emission direction by probe electrons at different accelerating voltages.}
\label{fig:energy}
\end{figure}

Our simulations indicate that due to the Pauli repulsion between their electron clouds, the impacted Si atom moving along the [110] lattice direction glances off either a neighbouring Si atom (Fig.~\ref{fig:trajectory}b,c) or the impurity (Fig.~\ref{fig:trajectory}h,i), which results in the change of its direction of motion and significant loss of kinetic energy. This loss can be reduced if the initial velocity of the impacted Si is directed slightly away from the neighbouring lattice site. Indeed, our simulations show that for the case of diagonal Bi jumps, a change of the emission angle $\Theta$ from 0 to 15{\textdegree} (see inset of Fig.~\ref{fig:energy}) results in a significant decrease of the threshold energy for indirect exchange---from 13.6 to 9.4\,eV. It further drops to 8.3\,eV for an emission angle of 30{\textdegree}, but remains the same for 45{\textdegree}. Similar behavior is also observed for lateral jumps, with differences between 0--0.6\,eV between the two directions and for the Sb dopant (see Table~\ref{tab:thresholds}).

\begin{table}[t]
\centering
\caption{\label{tab:thresholds} Calculated threshold energies (in eV) for the diagonal and lateral jumps of the Bi and Sb dopants as a function of the emission angle of the impacted Si atom.}
\begin{tabular}{lllllll}
$\Theta$ (\textdegree) & 0 & 5 & 10 & 15 & 30 & 45\\
\hline
Bi (diag.) & 13.6 & 11.6 & 10.3 & 9.4 & 8.3 & 8.3 \\
Bi (lat.) & 13.9 & 11.8 & 10.3 & 9.4 & 8.2 & 8.7 \\\hline
Sb (diag.) & 13.0 & 11.2 & 10.1 & 9.3 & 8.3 & 8.3 \\
Sb (lat.) & 13.6 & 11.6 & 10.3 & 9.4 & 8.2 & 8.6 \\
\hline
\end{tabular}
\end{table}

However, increase of the emission angle $\Theta$ also implies that the electron scattering angle is reduced~\cite{egerton_beam-induced_2013}, $\theta = \pi - 2\Theta$, and therefore the amount of kinetic energy transferred to the Si by the probe electron becomes smaller. In Fig.~\ref{fig:energy} we compare calculated threshold energies of Bi and Sb dopants averaged between diagonal and lateral directions for different emission angles (points) with the maximum kinetic energy that can be transferred to a stationary Si atom in the corresponding direction of motion at different electron energies (solid lines). For small emission angles, the decrease of the threshold energy is significantly faster than the decrease in the transferred energy. This suggest that an electron energies around 120\,keV should be sufficient for manipulation, and might present a useful future trade-off between speed and control. The direction of the energy transfer and thus manipulation probability could potentially also be controlled by tilting the beam~\cite{su_engineering_2019}.

We should also address the question of surface sputtering. Our simulations for the $(1\times1)$ Si surface reconstruction show that the threshold energy for sputtering Si atoms in the direction normal to the surface is only $\sim$8.5\,eV. We note, however, that surface sputtering depends on the surface reconstruction, oxidation and contamination, and therefore this value is merely an idealized estimate. Surface sputtering can be reduced by decreasing the electron energy, but it seems difficult to completely avoid this process for energies at which efficient manipulation can be performed.

\section*{Conclusions}
Solid-state quantum computers using nuclear spin qubits in silicon require the fabrication of a large ordered arrays of precisely positioned single donor atoms with a spacing of tens of nm, buried tens of nm below the crystal surface. This is a daunting challenge for traditional ion implantation and lithography techniques and remains the greatest hurdle for further progress. Deterministic ion implantation has been proposed as a possible solution~\cite{jamieson_controlled_2005}, but the achievable positional precision seems limited~\cite{pacheco_ion_2017}. Hydrogen-resist lithography in combination with scanning tunneling microscopy can achieve the required precision~\cite{fuechsle_single-atom_2012}, but qubit fabrication is complex and the choice of donor species is limited. Direct electron-beam manipulation of donor elements with atomic precision within the crystal lattice is thus a highly promising alternative. With process optimization both in terms of sample synthesis and tuning the primary beam energy, the manipulation of Sb dopants that we demostrate here may open a path to the deterministic and potentially scalable fabrication of solid-state nuclear spin qubits in silicon.

\begin{acknowledgement}

The authors thank J. Meyers for the cross-section specimen preparation and J. Kotakoski for useful discussions.

A.M., J.M., and T.S. were supported by the European Research Council (ERC) under the European Union’s Horizon 2020 research and innovation programme (Grant agreement No.~756277-ATMEN), and acknowledge computational resources provided by the Vienna Scientific Cluster. Electron microscopy work (B.M.H. and A.R.L.) was supported by the U.S. Department of Energy, Office of Science, Basic Energy Sciences, Division of Materials Sciences and Engineering, and sample growth (P.C.S. and J.S.) by the Laboratory Directed Research and Development Program of Oak Ridge National Laboratory, managed by UT-Battelle, LLC for the U.S. Department of Energy, and performed at the Oak Ridge National Laboratory’s Center for Nanophase Materials Sciences (CNMS), a U.S. Department of Energy Office of Science User Facility.

\end{acknowledgement}

\begin{suppinfo}

Supporting Information accompanying the article include a screen capture movie file of the Sb manipulation sequence shown in Fig.~\ref{fig:manipulation}, and a rendered video illustration of the indirect exchange mechanism following Fig.~\ref{fig:trajectory}. Open data of the simulated molecular dynamics trajectories, thermal MD trajectories, and relaxed split vacancy atomic configurations, as well as raw experimental scanning transmission electron microscopy images shown in Fig.~\ref{fig:interstitial} can be found on the University of Vienna institutional repository \href{https://doi.org/10.25365/phaidra.275}{Phaidra (doi:10.25365/phaidra.275)}. Open code to reproduce the image simulations in Fig.~\ref{fig:interstitial} can be found on \href{https://github.com/jacobjma/abTEM/tree/master/articles/2021_JPCC_manipulation_in_Si}{GitHub}.

\end{suppinfo}

\providecommand{\latin}[1]{#1}
\makeatletter
\providecommand{\doi}
  {\begingroup\let\do\@makeother\dospecials
  \catcode`\{=1 \catcode`\}=2 \doi@aux}
\providecommand{\doi@aux}[1]{\endgroup\texttt{#1}}
\makeatother
\providecommand*\mcitethebibliography{\thebibliography}
\csname @ifundefined\endcsname{endmcitethebibliography}
  {\let\endmcitethebibliography\endthebibliography}{}

\end{document}